\documentclass[aps,pra,reprint,longbibliography]{revtex4-1}
\usepackage{graphicx,amsmath, amssymb, bm, epstopdf,dsfont,float,comment}

\begin{document}

\title{Measurement of pure states of light in the orbital-angular-momentum basis using nine multipixel image acquisitions}

\author{Girish Kulkarni$^{\dagger}$, Suman Karan, and Anand K. Jha}
\email{akjha9@gmail.com;$^\dagger$ Presently at University of Ottawa, Canada}
\affiliation{Department of Physics, Indian Institute of Technology Kanpur, Kanpur, UP 208016, India\\}

\date{\today}

\begin{abstract}
The existing techniques for measuring high-dimensional pure states of light in the orbital angular momentum (OAM) basis either involve a large number of single-pixel data acquisitions and substantial postselection errors that increase with dimensionality, or involve substantial loss, or require interference with a reference beam of known phase. Here, we propose an interferometric technique that can measure an unknown pure state using only nine multipixel image acquisitions without involving postselection, loss, or a separate reference beam. The technique essentially measures two complex correlation functions of the input field and then employs a novel recursive postprocessing algorithm to infer the state. We experimentally demonstrate the technique for pure states up to dimensionality of 25, reporting a mean fidelity greater than 90 \% up to 11 dimensions. Our technique can significantly improve the performance of OAM-based information processing applications. 
\end{abstract}

\maketitle
\section{Introduction}
The orbital angular momentum (OAM) degree of freedom of light provides a discrete and unbounded basis for realizing high-dimensional classical and quantum states \cite{allen1992pra,erhard2018lsa}. In comparison to two-dimensional polarization states of light, high-dimensional states realized in the OAM basis can provide distinct advantages such as improved security and transmission bandwidth in quantum communication protocols \cite{karimipour2002pra,fujiwara2003prl}, enhanced system capacities and spectral efficiencies in classical communication protocols \cite{wang2012natphot,bozinovic2013science}, efficient gate implementations in computing schemes \cite{ralph2007pra,lanyon2009natphy}, enhanced sensitivity in metrology applications \cite{jha2011pra2,dambrosio2013natcomm}, and improved noise tolerance for fundamental tests of quantum theory \cite{kaszlikowski2000prl,collins2002prl}. 

In order to exploit such advantages of high-dimensional states, it is essential to develop robust and efficient techniques for measuring states of light in the OAM basis. In particular, the measurement of pure states of light has attracted significant interest \cite{nicolas2015njp,bent2015prx,malik2014natcomm,zhou2017lsa,zhao2017ol,derrico2017optica,singh2015pra}. The standard tomographic techniques based on mutually unbiased bases (MUBs) and symmetric informationally-complete positive operator-valued measures (SIC POVMs) have been experimentally limited by an increasing number of data acquisitions and post-processing times as the dimensionality of the input state increases \cite{nicolas2015njp,bent2015prx}. Furthermore, as these techniques rely on phase-flattening using computer-generated holograms followed by fiber-based detections, there are substantial postselection errors introduced due to the non-uniform fiber-coupling efficiencies experienced by different basis states \cite{qassim2014josab}. Another technique employs direct measurement, wherein the input state is inferred through sequential weak and strong measurements in the OAM and angle bases, respectively \cite{malik2014natcomm}. While this technique can measure a 27-dimensional state without much post-processing, its implementation involves custom-made refractive elements and multiple spatial light modulators (SLMs) making it cumbersome and lossy. Other existing techniques based on the rotational Doppler effect \cite{zhou2017lsa} and phase interferometry \cite{zhao2017ol,derrico2017optica,singh2015pra} have the drawback of requiring interference with a separate reference beam of known transverse phase. Thus, at present, there is a need to develop an efficient technique that can measure an unknown pure state without much loss and without requiring a separate reference beam of known transverse phase that is mutually coherent with the input field.
\begin{figure*}[t!]
\centering
\includegraphics[width=170mm,keepaspectratio]{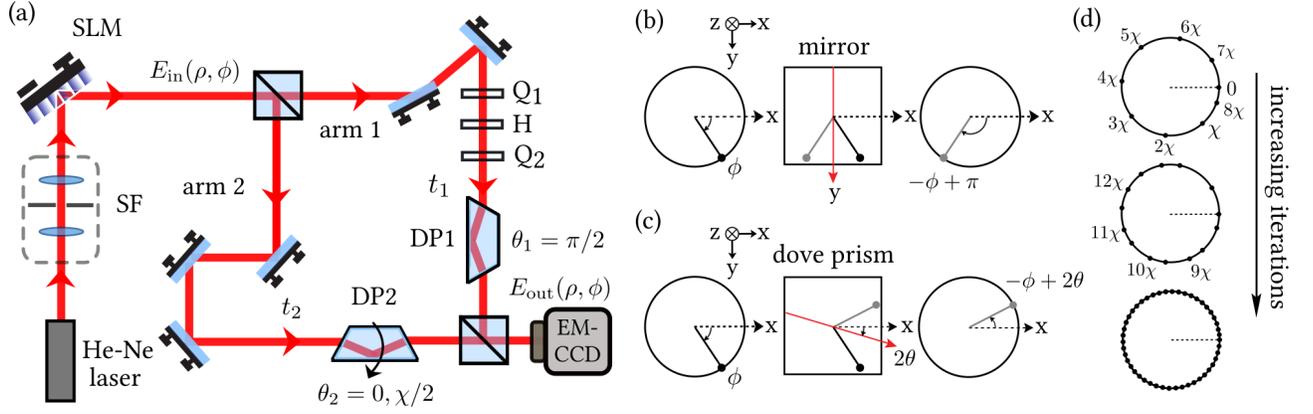}\vspace{-1mm}
\caption{Description of the experimental setup and technique: (a) A pure state of light described by Eq.~(\ref{pure-state}) is generated using a spatial light modulator (SLM), and is incident into a Mach-Zehnder interferometer. (b) A mirror reflects a beam about the y-axis transforming $\phi\to-\phi+\pi$. \textbf{c.} A dove prism (DP) rotated by angle $\theta$ reflects the beam about an axis rotated from the x-axis by an angle $2\theta$, effectively transforming $\phi\to-\phi+2\theta$. \textbf{d.} Schematically depicting how the complex input field $E_{\rm in}(\rho,\phi)$ is recursively inferred using Eq.~(\ref{recursive-relation}) with increasing number of iterations from top to bottom. When $\chi$ is an  irrational multiple of $\pi$, the recursive process eventually yields $E_{\rm in}(\rho,\phi)$ at a dense set of angles that uniformly cover the whole range of $0$ to $2\pi$, thereby allowing us to infer the input state. SF: spatial filter, Q$_1$, Q$_2$: quarter-wave plates, H: half-wave plate.} \vspace{-1mm}
\label{fig1}
\end{figure*}

Recently, in the context of measuring light beams that are \textit{incoherent} mixtures of various OAM basis states, a Mach-Zehnder interferometer with an odd and even number of mirrors has been demonstrated to be capable of measuring the OAM spectrum of the input field in a single-shot manner \cite{kulkarni2017natcomm}. The key feature of this interferometer is that it encodes the angular coherence function of the input field in the output interferogram. As a result, the OAM spectrum can be inferred by directly reading out the angular coherence function from an image of the output interferogram and performing an inverse Fourier transform. In this Letter, we adapt this interferometer to realize a novel technique for measuring pure OAM states or \textit{perfectly coherent} superpositions of different OAM basis states using nine intensity image acquisitions. Specifically, the technique uses one image of the bare input field to compute the transverse magnitude profile, and eight images of suitably-chosen output interferograms which through a recursive postprocessing algorithm yield the transverse phase profile of the field. The input state can then be inferred from the angular coherence function of this complex field. In comparison to tomographic techniques which have relied on sequential single-pixel acquisitions \cite{bent2015prx}, our technique is practically more efficient as the multipixel acquisitions effectively parallelize the information retrieval. Also, in comparison to direct measurement \cite{malik2014natcomm}, our technique does not involve any diffractive elements and therefore has negligible loss. Moreover, our technique involves no postselection and does not require a separate reference beam. Here, we experimentally demonstrate the technique using laboratory-generated pure states upto a dimensionality of 25, reporting a mean fidelity greater than 90 \% upto 11 dimensions.
\section{Theory}
We recall that the OAM basis states are the Laguerre-Gauss (LG) modes, represented as $LG^{l}_{p}(\rho,\phi)=LG^{l}_{p}(\rho)e^{il\phi}$ in the transverse polar co-ordinates \cite{allen1992pra}. The azimuthal index $l$ quantifies the OAM per photon in units of $\hbar$, and the radial index $p$ denotes the number of additional rings in the transverse profile of the mode. We consider a light beam that is a perfectly coherent superposition of LG modes with OAM indices ranging from $l=-N$ to $+N$, but with the same radial index $p$. While in principle $p$ can be any non-negative integer, for simplicity we assume $p=0$. The electric field $E_{\rm in}(\rho,\phi)$ at the waist plane of such a beam is written as
\begin{equation}\label{pure-state}
 E_{\rm in}(\rho,\phi)=\sum_{l=-N}^{+N} \alpha_{l} LG^{l}_{p=0}(\rho)e^{il\phi},
\end{equation}
where $\alpha_{l}$ are complex coefficients that satisfy $\sum_{l=-N}^{+N}|\alpha_{l}|^2=1$. For such a field, the correlation function $W(\rho,\phi_{1};\rho,\phi_{2})\equiv \langle E_{\rm in}(\rho,\phi_{1})E^*_{\rm in}(\rho,\phi_{2})\rangle$, where $\langle \cdots \rangle$ denotes an ensemble average, takes the factorized form \cite{mandel1995}
\begin{equation}\label{corrfunc-def}
 W(\rho,\phi_{1};\rho,\phi_{2})=E_{\rm in}(\rho,\phi_{1})E^*_{\rm in}(\rho,\phi_{2}).
\end{equation}
We emphasize that the above factorized form is valid only for perfectly coherent fields or pure states of light. 

We now consider the setup shown in Fig. 1a, where a beam of the kind described above is prepared using a spatial light modulator (SLM), and is incident into an interferometer. As depicted in Fig. 1b, the azimuthal co-ordinate $\phi$ of the LG modes is defined at the SLM with $\phi=0$ to be along the x-axis, and increases in the clockwise sense when looking into the beam propagation axis or z-axis. The rotation angle $\theta_{1(2)}$ of dove prism DP1(2) in arm 1(2) is defined such that $\theta_{1(2)}=0$ when its base is parallel to the xz plane, and positive rotation is in the clockwise sense when looking into the z-axis. As depicted in Figs. 1b and 1c respectively, each mirror can then be seen to transform $\phi\to-\phi+\pi$, and each dove prism can be seen to transform $\phi\to -\phi+2\theta_{1(2)}$. The output electric field $E_{\rm out}(\rho,\phi)$ at the camera is given by
\begin{multline}
 E_{\rm out}(\rho,\phi)=\sqrt{k_{1}}E_{\rm in}(\rho,\phi+\pi-2\theta_{1})e^{i(\omega_{0}t_{1}+\beta_{1})}\\+\sqrt{k_{2}}E_{\rm in}(\rho,-\phi+2\theta_{2})e^{i(\omega_{0}t_{2}+\beta_{2}+\tilde{\gamma})},
\end{multline}
where $\omega_{0}$ is the frequency, $k_{1}$ and $k_{2}$ are scaling factors involving beam-splitter ratios, etc., $t_{1(2)}$ and $\beta_{1(2)}$ correspond to the travel-time and non-dynamical phase, respectively, in arm 1(2), and $\tilde{\gamma}$ is a stochastic phase which incorporates the loss of temporal coherence. We have assumed the arm lengths to be much smaller than the Rayleigh range, and consequently the variation of the field's transverse profile due to spatial propagation can be ignored. We fix $\theta_{1}=\pi/2$, and keep it constant throughout the experiment. The output intensity $I_{\rm out}(\rho,\phi;\theta_{2},\delta)\equiv I_{\rm noise}(\rho,\phi;\theta_{2})+\langle E_{\rm out}(\rho,\phi)E^*_{\rm out}(\rho,\phi)\rangle$ can then be written as
\begin{multline}
 I_{\rm out}(\rho,\phi;\theta_{2},\delta)=I_{\rm noise}(\rho,\phi;\theta_{2})+k_{1}W(\rho,\phi;\rho,\phi)\\+k_{2}W(\rho,-\phi+2\theta_{2};\rho,-\phi+2\theta_{2})\\+\gamma\sqrt{k_{1}k_{2}}W(\rho,\phi;\rho,-\phi+2\theta_{2})e^{i\delta}+\mathrm{c.c}.
\end{multline}
Here, $\delta$ is the total phase difference between the two arms given by $\delta\equiv[\omega_{0}(t_{1}-t_{2})+\beta_{1}-\beta_{2}+\mathrm{arg}(\langle e^{-i\tilde{\gamma}} \rangle)]$ modulo $2\pi$, and $\gamma=|\langle e^{-i\tilde{\gamma}} \rangle|$ is the degree of temporal coherence. The term $I_{\rm noise}(\rho,\phi;\theta_{2})$ contains noise due to ambient light and intensity aberrations from dove prisms. As $\theta_{1}$ remains fixed, this noise term only depends on $\theta_{2}$.

We set $\theta_{2}=0$, and record four output intensity shots corresponding to $\delta=0,\pi/2,\pi$, and $3\pi/2$ to obtain  
\begin{subequations}\label{corrfunc1}
\begin{align}\notag
\mathrm{Re}\left[W(\rho,\phi;\rho,-\phi)\right]&=\frac{1}{4\gamma\sqrt{k_{1}k_{2}}}\Big\{I_{\rm out}(\rho,\phi;0,0)\Big.\\\Big.&\hspace{2mm}-I_{\rm out}(\rho,\phi;0,\pi)\Big\}, \\\notag
\mathrm{Im}\left[W(\rho,\phi;\rho,-\phi)\right]&=\frac{1}{4\gamma\sqrt{k_{1}k_{2}}}\Big\{I_{\rm out}(\rho,\phi;0,3\pi/2)\Big.\\\Big.&\hspace{0mm}-I_{\rm out}(\rho,\phi;0,\pi/2)\Big\}.
\end{align}
\end{subequations}

Next, we set $\theta_{2}=\chi/2$, where $\chi$ is a judiciously chosen value, and record four output intensity shots corresponding to $\delta=0,\pi/2,\pi,$ and $3\pi/2$ to obtain
\begin{subequations}\label{corrfunc2}
\begin{align}\notag
\mathrm{Re}\left[W(\rho,\phi;\rho,-\phi+\chi)\right]&=\frac{1}{4\gamma\sqrt{k_{1}k_{2}}}\Big\{I_{\rm out}(\rho,\phi;\chi/2,0)\Big.\\\Big.&\hspace{2mm}-I_{\rm out}(\rho,\phi;\chi/2,\pi)\Big\}, \\\notag
\mathrm{Im}\left[W(\rho,\phi;\rho,-\phi+\chi)\right]&=\frac{1}{4\gamma\sqrt{k_{1}k_{2}}}\Big\{I_{\rm out}(\rho,\phi;\chi/2,3\pi/2)\Big.\\\Big.&\hspace{2mm}-I_{\rm out}(\rho,\phi;\chi/2,\pi/2)\Big\}.
\end{align}
\end{subequations}
As the noise term $I_{\rm noise}(\rho,\phi;\theta_{2})$ does not explicitly depend on $\delta$, it is canceled out from equations (\ref{corrfunc1}) and (\ref{corrfunc2}). Moreover, a precise characterization of the parameters $\gamma,k_{1},$ and $k_{2}$ is not required as they appear only as an overall scaling factor. Thus, the eight shots allow us to infer the complex correlation functions $W(\rho,\phi;\rho,-\phi)$ and $W(\rho,\phi;\rho,-\phi+\chi)$ in a noise-insensitive manner. 
\begin{figure*}[t!]
\centering
\includegraphics[width=170mm,keepaspectratio]{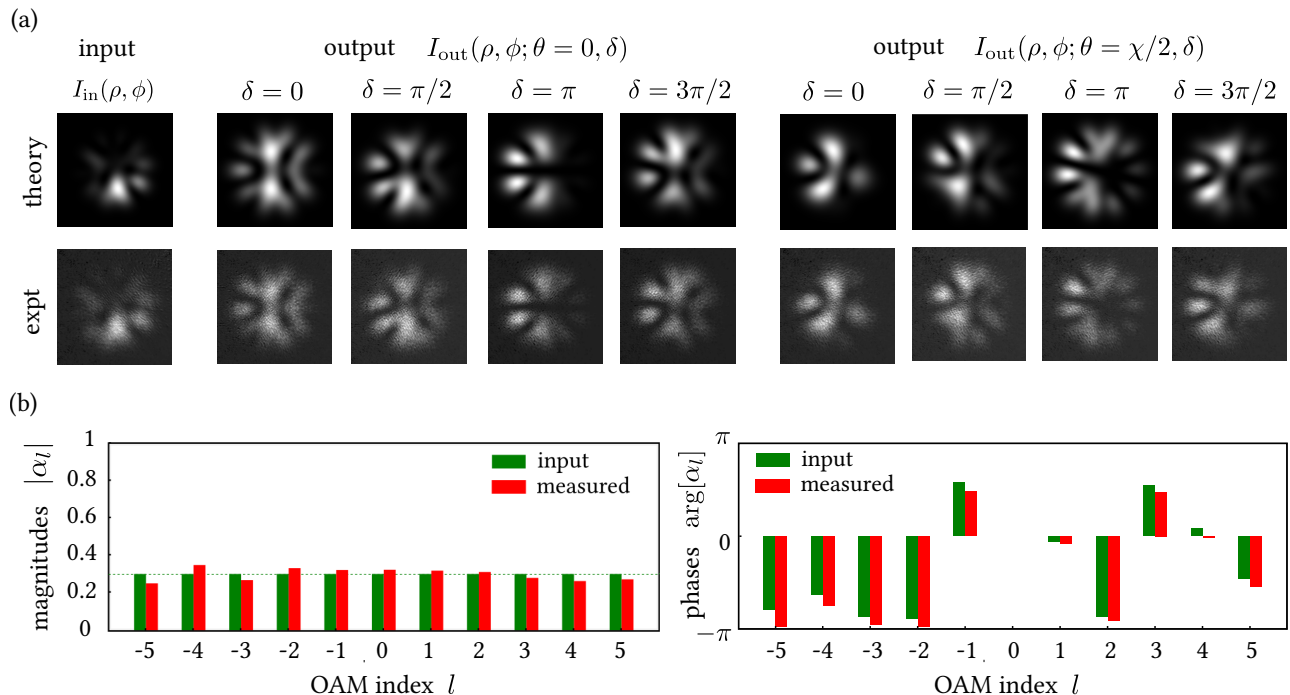}\vspace{0mm}
\caption{Experimental measurement of an 11-dimensional state $|\psi_{\rm in}\rangle=\sum_{l=-5}^{+5}\alpha_{l}|l\rangle$. (a) Experimental and theoretical images of the input intensity and output interferograms at $\delta=0, \pi/2, \pi, 3\pi/2$ for $\theta_{2}=0$ and $\theta_{2}=\chi/2$. (b) Bar plots depicting the measured magnitudes and phases of the coefficients $\alpha_{l}$ of the input state. As the global phase is arbitrary, we have chosen it such that $\mathrm{arg}[\alpha_{0}]=0$ for the input and measured states. The fidelity of our measurement is $F=|\langle \psi_{\rm in}|\psi_{\rm meas}\rangle|^2=96.78 \%$.} 
\label{fig1}
\end{figure*}

We will now show that the correlation functions $W(\rho,\phi;\rho,-\phi)$ and $W(\rho,\phi;\rho,-\phi+\chi)$ are in principle sufficient to completely infer the input field $E_{\rm in}(\rho,\phi)$ in a recursive manner. We note that upon dividing the two correlation functions and using Eq.~(\ref{corrfunc-def}), we obtain 
\begin{align}\label{recursive-relation}
E_{\rm in}(\rho,-\phi+\chi)=\frac{W^*(\rho,\phi;\rho,-\phi+\chi)}{W^*(\rho,\phi;\rho,-\phi)}E_{\rm in}(\rho,-\phi).
\end{align}
We begin the recursive process starting from $E_{\rm in}(\rho,\phi=0)$, which can be computed from $W(\rho,\phi;\rho,-\phi)$ as $E_{\rm in}(\rho,\phi=0)=\sqrt{W(\rho,\phi=0;\rho,-\phi=0)}$. As the global phase of the field is arbitrary, we can assume without loss of generality that the field at $\phi=0$ is real, i.e, $\mathrm{arg}[E_{\rm in}(\rho,\phi=0)]=0$. Writing Eq.~(\ref{recursive-relation}) for $\phi=0$, we can obtain $E_{\rm in}(\rho,\phi=\chi)$ from $E_{\rm in}(\rho,\phi=0)$. Then, writing Eq.~(\ref{recursive-relation}) for $\phi=-\chi$, we can obtain $E_{\rm in}(\rho,\phi=2\chi)$, and so on. Now if $\chi$ is not a rational multiple of $\pi$, then as shown in Fig. 1d the iterations wrap around without falling back on a previously sampled angular location. After a large number of iterations, the set of angular locations $\phi=\{0,\chi,2\chi,...\}$ modulo $2\pi$ uniformly covers the whole range from $0$ to $2\pi$. In this way, the complex electric field $E_{\rm in}(\rho,\phi)$ can in principle be completely inferred, and by inverting Eq.~(\ref{pure-state}) the complex coefficients $\alpha_{l}$ of the pure state can be computed. Thus, the input pure state can in principle be measured using only eight multipixel intensity acquisitions irrespective of the dimensionality of the input state. 

In practice, however, the above process leads to highly noisy measurements of input states. This is primarily because the division in Eq.~(\ref{recursive-relation}) diverges when the denominator $W^*(\rho,\phi;\rho,-\phi)$ takes small values, leading to amplification of small errors through the subsequent recursive iterations. However, this problem can be circumvented in the following manner. We use the phase argument of Eq.~(\ref{recursive-relation}) given by
\begin{multline}\label{recursive-relation-phase}
\mathrm{arg}\left[E_{\rm in}(\rho,-\phi+\chi)\right]=\mathrm{arg}\left[W^*(\rho,\phi;\rho,-\phi+\chi)\right]\\-\mathrm{arg}\left[W^*(\rho,\phi;\rho,-\phi)\right]+\mathrm{arg}\left[E_{\rm in}(\rho,-\phi)\right],
\end{multline}
to recursively infer only the phase $\mathrm{arg}[E_{\rm in}(\rho,\phi)]$ of the input field. In contrast with Eq.~(\ref{recursive-relation}) which involves division and multiplication in each iteration, the above recursive relation (\ref{recursive-relation-phase}) only involves subtraction and addition. As a result, small errors do not get amplified through the recursive process, and the phase $\mathrm{arg}[E_{\rm in}(\rho,\phi)]$ is obtained in a noise-insensitive manner. The magnitude $|E_{\rm in}(\rho,\phi)|$ can be separately inferred by acquiring an intensity acquisition $I_{\rm in}(\rho,\phi)$ of the input field, and computing $|E_{\rm in}(\rho,\phi)|=\sqrt{I_{\rm in}(\rho,\phi)}$. Thus, the measurement procedure now involves nine multipixel intensity acquisitions -- eight shots corresponding to eight interferograms which recursively yield the phase profile $\mathrm{arg}[E_{\rm in}(\rho,\phi)]$, and a ninth shot of the bare input field which yields the magnitude profile $|E_{\rm in}(\rho,\phi)|$. In this way, the complex input field is completely determined. 

The final step is to compute the complex coefficients of the input state from the complex field. This can in principle be done by simply inverting Eq.~(\ref{pure-state}) to get $\alpha_{l}=\iint E_{\rm in}(\rho,\phi) LG^{l}_{p=0}(\rho)e^{-il\phi}\,\rho\,\mathrm{d}\rho\,\mathrm{d}\phi$. However, we have numerically observed that this direct overlap integral is more sensitive to noise due to pixelation in $E_{\rm in}(\rho,\phi)$ and errors in the beam waist of the $LG^{l}_{p=0}(\rho)$ function. Therefore, we first compute the angular coherence function $\bar{W}(\phi_{1},\phi_{2})$ of the field as \cite{jha2011pra,kulkarni2017natcomm}
\begin{equation}
 \bar{W}(\phi_{1},\phi_{2})=\int_{0}^{\infty} E_{\rm in}(\rho,\phi_{1})E^*_{\rm in}(\rho,\phi_{2})\,\rho\,\mathrm{d}\rho,
\end{equation}
and then use equations (\ref{pure-state}) and (\ref{corrfunc-def}) to compute the measured OAM cross-correlation matrix 
\begin{equation}\label{oam-matrix}
 \alpha_{l_{1}}\alpha^*_{l_{2}}=\frac{\int_{0}^{2\pi}\int_{0}^{2\pi} \bar{W}(\phi_{1},\phi_{2})\,e^{-i(l_{1}\phi_{1}-l_{2}\phi_{2})}\mathrm{d}\phi_{1}\mathrm{d}\phi_{2}}{\int_{0}^{\infty} LG^{l_{1}}_{p=0}(\rho) LG^{*l_{2}}_{p=0}(\rho)\,\rho\,\mathrm{d}\rho}.
\end{equation}
The coefficients $\alpha_{l}$ are the elements of the normalized principal eigenvector of the above matrix. In comparison to the direct overlap integral method, we find that the radial integrations in equations (9) and (10) reduce the pixelation-related and beam waist-related noise. 
\section{Experimental Results}
As shown in Fig 1a, a 5 mW horizontally-polarized He-Ne laser beam is spatially filtered and made incident on a Holoeye Pluto spatial light modulator (SLM). A hologram of 1000x1000 pixels is prepared using the method by Arrizon \textit{et al.} \cite{arrizon2007josaa} to produce a pure state field described by Eq.~(\ref{pure-state}), which is then made incident into the Mach-Zehnder interferometer. The quarter-wave plates Q$_{1}$ and Q$_{2}$ are oriented at 45 degrees with respect to the horizontal direction. When the half-wave plate H is rotated clockwise by a certain angle, a geometric phase-shift of twice that angle is introduced between the two arms \cite{jha2008prl}. A box was used to cover the interferometer to reduce ambient air movements, and thereby stabilize the relative phase to within a few degrees for about 1 minute. The conditions of $\delta\approx0,\pi/2,\pi,3\pi/2$ can be identified by displaying a hologram corresponding to a Gaussian beam and measuring the total output intensity while manually rotating the half-wave plate using a long probe without opening the box. When the desired condition of $\delta$ is realized, the hologram is switched to the input state within 1s and the corresponding interferogram is acquired using an Andor EMCCD camera. 

First, we acquire the intensity profile $I_{\rm in}(\rho,\phi)$ of the bare input field. Next, for $\theta_{2}=0$ we record the output interferograms $I_{\rm out}(\rho,\phi;\theta_{2}=0,\delta)$ for $\delta=0,\pi/2,\pi,$ and $3\pi/2$. Subsequently, we set $\theta_{2}=\chi/2=14.5$ degrees and record the output interferograms $I_{\rm out}(\rho,\phi;\theta_{2}=\chi/2,\delta)$ for $\delta=0,\pi/2,\pi,$ and $3\pi/2$. The Dove prism rotation mounts had a least count of $0.5$ degrees. We chose $\chi/2=14.5$ because $\chi=29$ is coprime with 360 ensuring that $n\chi$, where $n=1,2,3,...,360$, uniformly sample the entire angular space from $0$ to $360$ degrees with a sampling width of $1$ degree. We have ignored the small change in the polarization vector of the field induced by the Dove prism rotation \cite{padgett1999jmo,moreno2003optcomm}. The output 512 x 512 image arrays in Cartesian co-ordinates $x$ and $y$ from the camera were cleaned using a low-pass computational Fourier filter, and converted to 200 x 360 arrays in polar co-ordinates $\rho$ and $\phi$ using the polarTransform module in Python. As the input state is Hermitian, in principle $W(\rho,\phi_{1};\rho,\phi_{2})=W^*(\rho,\phi_{2};\rho,\phi_{1})$. In practice, the experimentally measured functions $W(\rho,\phi;\rho,-\phi)$ and $W(\rho,\phi;\rho,-\phi+\chi)$ may be non-Hermitian due to the presence of noise and experimental errors in $\delta$ for the various acquisitions. The effect of such errors can be reduced by retaining only the Hermitian parts of $W(\rho,\phi;\rho,-\phi)$ and $W(\rho,\phi;\rho,-\phi+\chi)$ for the recursive procedure. Also, for evaluating the matrix elements in Eq.~(\ref{oam-matrix}), the beam waist parameter for the $LG^{l}_{p=0}(\rho)$ functions can be chosen approximately such that the mode with the largest OAM index $l=N$ fits the radial extent of the input beam. We have verified that the measurement procedure is robust to minor variations in the beam waist choice. We also note that the arms of our interferometer, each about 0.3 m long, were much smaller than the Rayleigh range of our input beam, which was about 30 m. As a result, the Gouy phases acquired by the constituent OAM modes were negligibly small and could be ignored \cite{padgett1995optcomm}. 

In Fig 2a, we depict the theoretical and experimental intensity profiles corresponding to these nine acquisitions for an 11-dimensional input state whose coefficients $\alpha_{l}$ have uniform magnitudes $|\alpha_{l}|=1/\sqrt{11}$ and phases $\mathrm{arg}[\alpha_{l}]$ randomly chosen in the interval $[-\pi,\pi]$. While our technique can be used for measuring an arbitrary pure state, we have deliberately chosen to present our results for a pure state whose coefficients have uniform magnitudes for two reasons: one, to ensure that the state does not have a highly skewed support in the available dimensions; and two, to make it easier to visually compare the match between the input and measured coefficients. Following the procedure described in Sec. II, we compute the measured coefficients and depict their magnitudes and phases in Fig 2b. We find an excellent match between the input and measured coefficients. As the global phases of the input and measured states are arbitrary, we have chosen them such that $\mathrm{arg}[\alpha_{0}]=0$. If we denote the input state as $|\psi_{\rm in}\rangle=\sum_{l=-N}^{+N}\alpha_{l}|l\rangle$ and the measured state as $|\psi_{\rm meas}\rangle=\sum_{l=-N}^{+N}\bar{\alpha}_{l}|l\rangle$, we can compute the fidelity $F$ as defined in Ref.~\cite{bengtsson2017book} to find $F=|\langle \psi_{\rm in}|\psi_{\rm meas}\rangle|^2=|\sum_{l=-N}^{N}\alpha^*_{l}\bar{\alpha}_{l}|^2=96.78\%$.
\begin{figure}[t!]
\centering
\includegraphics[width=80mm,keepaspectratio]{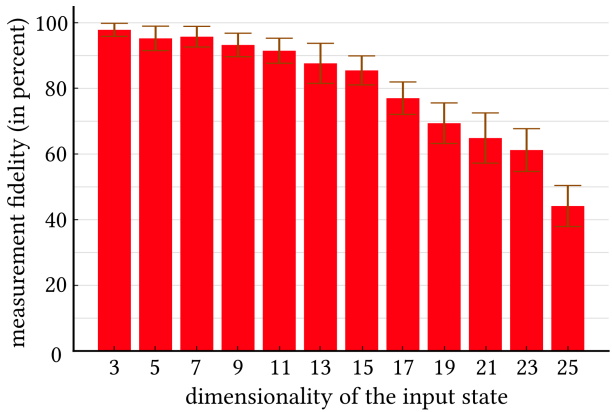}\vspace{-2mm}
\caption{Bar graph depicting how the experimentally measured fidelity varies with respect to dimensionality $(2N+1)$ of input states $|\psi_{\rm in}\rangle=\sum_{l=-N}^{+N}\alpha_{l}|l\rangle$, whose complex coefficients $\alpha_{l}$ have uniform magnitudes and random phases.} \vspace{-2mm}
\label{fig1}
\end{figure}

In order to quantify the scaling of the fidelity with respect to dimensionality in our setup, we performed experiments for several input states described by Eq.~(\ref{pure-state}) with increasing dimensionality $(2N+1)$, where coefficients $\alpha_{l}$ were chosen to have uniform magnitudes and random phases. While the technique is in principle capable of measuring pure states of arbitrarily high dimensionalities, as depicted in Fig. 3 we experimentally find a gradual reduction in the mean fidelity with increasing dimensionality. Moreover, the variation in the fidelity is higher for input states of higher dimensionalities. We attribute this behavior mainly to errors in the input state generation from the SLM, errors in the dove prism rotation angle and geometric phase shifts using waveplates, and the finite pixelation of the camera sensor. As may be noted from Fig 3, our setup was able to measure states with mean fidelity greater than 90 \% upto a dimensionality of 11. We expect that with better experimental controls, the technique can be scaled up to measure states of larger dimensionalities with higher fidelities. 
\section{Conclusions}
In summary, we have proposed and experimentally demonstrated an efficient technique for measuring pure states of light in the OAM basis. In comparison to previous tomographic techniques which rely on a large number of sequential single-pixel acquisitions \cite{bent2015prx}, our technique involves acquiring only nine images using a multipixel camera and postprocessing them with a novel recursive algorithm to infer the input state. Moreover, in contrast with the direct measurement technique \cite{malik2014natcomm}, our technique does not involve diffractive elements and therefore has negligible loss. Furthermore, our technique does not involve postselection, does not require a separate mutually coherent reference beam, and is robust to ambient intensity noise and errors in the characterization of setup parameters. However, a drawback of our technique is that for measuring single-photon fields, it is essential to stabilize the interferometer for long acquisition times which can be challenging. 

In future, we expect this technique to lead to improvements in the system capacities and transmission bandwidths of communication protocols exploiting OAM states of light by practically enabling measurement of states with higher dimensionalities. We also expect that this technique can be suitably adapted to measure pure states of light in other spatial bases such as Hermite-Gauss and Ince-Gauss bases. Finally, this work may also eventually lead to novel techniques for measuring arbitrary mixed states of light.\\

\section*{Acknowledgments} We thank Shaurya Aarav for several insightful discussions. We acknowledge financial
support through grant no. EMR/2015/001931 from the Science and Engineering Research Board and through grant no. DST/ICPS/QuST/Theme -1/2019
from the Department of Science \& Technology, Government of India.


\end{document}